\documentclass[runningheads]{llncs}
\usepackage[T1]{fontenc}
\usepackage{graphicx}
\usepackage{amsmath,amssymb,amsfonts}
\usepackage{algorithmic}
\usepackage{textcomp}
\usepackage{listings}
\usepackage{pgfplots}
\usepackage{pgfplotstable}
\usepackage{subcaption}
\usepackage{csquotes}
\usepackage{multirow}
\usepackage{colortbl}
\usepackage{tabularx}
\usepackage{booktabs}
\usepackage{hyperref}
\usepackage[capitalise,noabbrev]{cleveref}
\usepackage{xcolor}
\usepackage[inline,shortlabels]{enumitem}
\usepackage[scaled=.98]{inconsolata}

\newcolumntype{Y}{>{\centering\arraybackslash}X}

\usetikzlibrary{patterns, patterns.meta}
\usepgfplotslibrary{groupplots}

\pgfdeclarepattern{
  name=hatch,
  parameters={\hatchsize,\hatchangle,\hatchlinewidth},
  bottom left={\pgfpoint{-.1pt}{-.1pt}},
  top right={\pgfpoint{\hatchsize+.1pt}{\hatchsize+.1pt}},
  tile size={\pgfpoint{\hatchsize}{\hatchsize}},
  tile transformation={\pgftransformrotate{\hatchangle}},
  code={
    \pgfsetlinewidth{\hatchlinewidth}
    \pgfpathmoveto{\pgfpoint{-.1pt}{-.1pt}}
    \pgfpathlineto{\pgfpoint{\hatchsize+.1pt}{\hatchsize+.1pt}}
    \pgfpathmoveto{\pgfpoint{-.1pt}{\hatchsize+.1pt}}
    \pgfpathlineto{\pgfpoint{\hatchsize+.1pt}{-.1pt}}
    \pgfusepath{stroke}
  }
}

\tikzset{
  hatch size/.store in=\hatchsize,
  hatch angle/.store in=\hatchangle,
  hatch line width/.store in=\hatchlinewidth,
  hatch size=7pt,
  hatch angle=0pt,
  hatch line width=.5pt,
}

\pgfplotscreateplotcyclelist{listOverheadAbs}{%
    {black, fill=white},
    {black, fill=gray},
    {black, fill=gray, postaction={opacity=0.8,pattern color=black,pattern=north east lines}},
    {black, fill=gray, postaction={pattern color=white!80!black,pattern=hatch}}
}

\pgfplotsset{compat=1.18}

\newenvironment{enumi}{\begin{enumerate*}[label=(\roman*)]}{\end{enumerate*}}

\newcommand{\NHRText}{%
The authors would like to thank the Federal Ministry of Research, Technology and Space
and the state governments (www.nhr-verein.de/unsere-partner) for supporting this
work as part of the joint funding of National High Performance Computing
(NHR).
}

\newcommand{\CPP}{C\nolinebreak\hspace{-.05em}\raisebox{.4ex}{\tiny\bf +}\nolinebreak\hspace{-.10em}\raisebox{.4ex}{\tiny\bf +}}

\lstdefinelanguage{llvm}{
  morecomment=[l]{;},
  morestring=[b]",
  sensitive=true,
  morekeywords={
    define, declare, global, constant, private, internal, linkonce, linkonce_odr, weak, weak_odr, appending,
    ret, br, switch, invoke, resume,
    add, fadd, sub, fsub, mul, fmul, udiv, sdiv, fdiv, urem, srem, frem,
    shl, lshr, ashr, and, or, xor,
    extractelement, insertelement, shufflevector,
    extractvalue, insertvalue,
    alloca, load, store, fence, cmpxchg, atomicrmw, getelementptr,
    trunc, zext, sext, fptrunc, fpext, fptoui, fptosi, uitofp, sitofp, ptrtoint, inttoptr, bitcast, addrspacecast,
    icmp, fcmp, phi, select, call, va_arg, landingpad, catchpad, cleanuppad,
    eq, ne, sgt, sge, slt, sle, ugt, uge, ult, ule
  },
  morekeywords=[2]{i1, i8, i16, i32, i64, float, double, void, label, metadata, ptr},
}

\begin{document}
\title{Extending Contract Verification for Parallel~Programming~Models to Fortran}
%
%
\author{Yussur Mustafa Oraji\inst{1}\orcidID{0009-0004-9922-3112} \and
Christian~Bischof\inst{1}\orcidID{0000-0003-2711-3032}}
\authorrunning{Y. Oraji et al.}
%
\institute{Department of Computer Science, TU Darmstadt University, Darmstadt, Germany \email{\{yussur.oraji,christian.bischof\}@tu-darmstadt.de}}
\maketitle              
\begin{abstract}
High-performance computing often relies on parallel programming models such as MPI for distributed-memory systems.
While powerful, these models are prone to subtle programming errors,
leading to development of multiple correctness checking tools.
However, these are often limited to C/\CPP\ codes, tied to specific library implementations, or restricted to certain error classes.
Building on our prior work with CoVer, a generic, contract-based verification framework for parallel programming models,
we extend CoVer's applicability to Fortran, enabling static and dynamic analysis across multiple programming languages.
We adapted language-specific contract definitions and modified the analyses to support both C/\CPP\ and Fortran programs.
Our evaluation demonstrates that the enhanced version preserves CoVer's analysis accuracy and even revealed a bug in the MPI-BugBench testing framework,
underscoring the effectiveness of the approach.
The Fortran port of CoVer turns out to be substantially more efficient than the state-of-the-art tool MUST, while maintaining generality across languages.

\keywords{MPI, Contracts, Correctness, Fortran, Static Analysis, Dynamic Analysis}
\end{abstract}

\section{Introduction}
\label{sec:introduction}

In order to perform large-scale computations on distributed-memory clusters,
the use of parallel programming models such as MPI \cite{messagepassinginterfaceforumMPIMessagePassingInterface2025},
OpenSHMEM \cite{openshmemcommitteeOpenSHMEMApplicationProgramming2020} or NCCL \cite{nvidiaNVIDIACollectiveCommunication2025} is necessary.
While they offer powerful programming paradigms, their use is typically very error-prone.
Programming errors may symptomize as program crashes, deadlocks, or silent data corruption,
any of which costs both time and compute resources to debug and correct.

To avoid this, correctness checking tools have been developed.
These range from the static, compile-time tools \cite{drosteMPIcheckerStaticAnalysis2015,burakSPMDIRUnifying2025},
to dynamic, runtime checkers \cite{hilbrichMUSTScalableApproach2010,schwitanskiRMASanitizerGeneralizedRuntime2024}.
However, most correctness checkers are limited to one or a set of known parallel programming models for analysis;
this is the case for all tools mentioned.

Previously, we presented CoVer \cite{orajiVerifyingMPIAPI2026}, a contract-based static analysis approach for verifying the use of parallel programming models.
Due to its contract-based nature, it is not bound to any one parallel programming model;
support can be extended arbitrarily using fitting contract annotations for any models' API functions.
To improve analysis accuracy, we later presented CoVer-Dynamic \cite{orajiDynamicContractAnalysis2026}, a dynamic analysis for the same contracts.
Thus, once the contract annotations are present, analysis can be performed using either static or dynamic analysis.
However, analysis using CoVer is currently limited to C and \CPP, with the notable exception of the still widely used Fortran \cite{lagunaLargescaleStudyMPI2019}.

Some tools do support Fortran, such as MUST/RMASanitizer \cite{hilbrichMUSTScalableApproach2010} and PARCOACH \cite{saillardPARCOACHCombiningStatic2014}, but their support is limited.
PARCOACH only supports the old \texttt{mpif.h} or \texttt{use mpi} bindings,
while MUST can perform analysis for the Fortran 2008 MPI module only when using a specially modified MPICH MPI implementation.
Finally, to our knowledge, there is currently no tool capable of error checking for multiple parallel programming models statically for Fortran.

This paper, then, presents an extension to the CoVer framework to work on Fortran code.
In the spirit of generality, our implementation remains independent of any one parallel programming model.
Our work spans both the static and dynamic modes of CoVer, allowing the use of either for the analysis of Fortran code.
We ensured support for both older and newer Fortran standards, and tested this using the older MPI bindings (\texttt{mpif.h}, \texttt{use mpi}),
as well as the newer ones (\texttt{use mpi\_f08}, \texttt{use mpi\_f08} + TS29113).

We evaluate the classification quality of our port using the Fortran version of MPI-BugBench \cite{orajiExtendingMPICorrectness2025} to check for accuracy discrepancies to the C baseline.
Further, we compare the overhead induced on equivalent C and Fortran versions of miniWeather \cite{normanMiniWeather2020} and the PRK Stencil Kernel \cite{vanderwijngaartParallelResearchKernels2014},
for which we wrote a new Fortran port, to check for performance regressions in the dynamic analysis.

Our work provided immediate results, detecting an invalid test case in MPI-BugBench.
We published a fix to the upstream MPI-BugBench project \cite{orajiFixUnmatchedWait2026}, which has since been merged.

\section{Background}
\label{sec:Background}

Implementing support for Fortran required changes within the static CoVer analysis (referred to as CoVer-Static hereafter),
as well as the instrumentation performed for CoVer-Dynamic.
This section introduces the underlying shared contract language, as well as the two modes of operation (static/dynamic),
whose adaption for Fortran code is presented in \cref{sec:implementation}.

\subsection{CoVer Contract Language}
\label{sec:contract_language}

The CoVer contract language \cite{orajiVerifyingMPIAPI2026} allows defining requirements,
which must hold before or after an API call invocation.
For example, the MPI standard requires that, before any other API calls occur,
a call to \texttt{MPI\_Init} is made (simplifying, ignoring the sessions model).
\begin{figure}[tbp]
\begin{lstlisting}
int MPI_Get(^*\dots*^) CONTRACT(
  PRE { call!(MPI_Init) }
  POST {
    no! (read!(*0)) until! (call_tag!(rma_complete,$:7)),
    no! (write!(*0)) until! (call_tag!(rma_complete,$:7))
});
int MPI_Win_fence(int assert, MPI_Win win) CONTRACT( TAGS { rma_complete(1) });
int MPI_Win_unlock_all(MPI_Win win) CONTRACT( TAGS { rma_complete(0) });
\end{lstlisting}
\caption{Example of contract definitions (C/\CPP).}
\label{fig:contr_def_examples_c}
\end{figure}
This requirement can be made concrete using contracts.
As an example, one such contract requiring a call to \texttt{MPI\_Init} before any call to \texttt{MPI\_Get}, is shown in \cref{fig:contr_def_examples_c}.
The declaration of \texttt{MPI\_Get} receives an additional \texttt{CONTRACT} annotation.
Within, it defines a requirement that must hold prior to any callsite of \texttt{MPI\_Get}, as indicated by the \texttt{PRE} \emph{scope}.
Finally, in that scope is an \emph{operation}, more specifically a call operation with \texttt{MPI\_Init} as the target.

Further, there are additional requirements in the postcondition scope (\texttt{POST}), which in this case forbid data races.
\texttt{MPI\_Get} is a non-blocking call, which means that the function may return prior to the communication being completed.
This being the case, any memory access to the supplied buffer would be erroneous,
as the resulting value in the buffer is dependent on whether the communication has completed yet or not, leading to a data race.
Instead, the buffer should \emph{not} be accessed \emph{until} a call to a corresponding completion routine is performed.
This semantic structure can be applied to the contract definition, as shown in \cref{fig:contr_def_examples_c},
which forbids reading/writing accesses to the memory pointed to by the first parameter of \texttt{MPI\_Get} (the buffer)
using the \texttt{read(*0)}/\texttt{write(*0)} operations.

Describing the completion is more complex, as multiple functions can complete the communication, such as \texttt{MPI\_Win\_fence} or \texttt{MPI\_Win\_unlock\_all},
and only if called using the same window handle.
To specify the releasing operation as a call to any of these functions, the \emph{tag} system is used.
Both \texttt{MPI\_Win\_fence} and \texttt{MPI\_Win\_unlock\_all} are given a shared tag, in this case called \texttt{rma\_complete}.
Then, by using a \emph{call-tag} operation instead of a normal call operation, the tag is specified as the target, allowing either function to release the requirement.
Finally, to map the window buffer of the \texttt{MPI\_Get} call to the one given in the completion call, we additionally specify the mapping parameter \texttt{\$:7}.
This implies that in order for the function call to match, the parameter given in the tag definition (which is the parameter of the window handle in the completion call)
must match the eighth parameter of the \texttt{MPI\_Get} call, which is its corresponding window handle.
Thus, the contract states that there may be no reading or writing operations to the buffer until the communication is completed.

\subsection{CoVer-Static}
\label{sec:cover-static}

CoVer-Static is a contract-based static analysis tool which uses a formal data-flow analysis \cite{orajiVerifyingMPIAPI2026},
implemented on top of LLVM \cite{LLVMCompilerInfrastructure} as a set of compile-time passes
for the LLVM Intermediate Representation (LLVM IR).

\begin{figure}[tbp]
    \centering
    \includegraphics[width=0.9\textwidth]{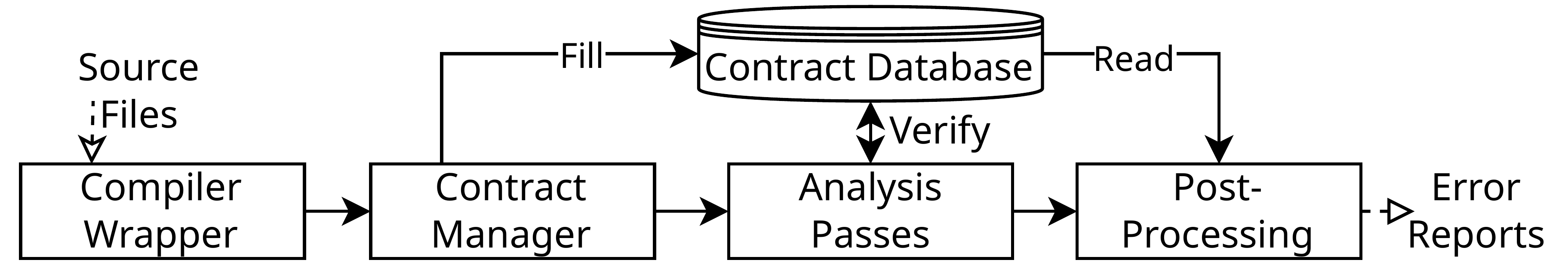}
    \caption{Architecture of CoVer-Static, from \cite{orajiVerifyingMPIAPI2026}.}
    \label{fig:arch_cover-static}
\end{figure}
CoVer-Static is split into multiple parts, as shown in \cref{fig:arch_cover-static}:
\begin{enumi}
    \item the compiler wrapper,
    \item the contract manager,
    \item the analysis passes and
    \item post-processing.
\end{enumi}
The compiler wrapper is used to perform whole-program analysis.
Instead of creating object files when compiling a file,
the wrapper instead emits LLVM IR.
When the program is supposed to be linked, the LLVM IR is linked together instead.
The resulting monolithic IR is then passed on to the analyses, after which it is compiled and linked.
The compiler wrapper can optionally include predefined contracts for MPI and OpenSHMEM as well.

Once it is time for analysis, the contract manager runs first.
The contract manager parses the string contract annotations into an internal representation,
which is then passed on to the analyses.

The analyses run on these internal representations.
Using the data-flow approach described in \cite{orajiVerifyingMPIAPI2026}, the analyses mark the operations within the contracts as fulfilled or violated.
The data-flow analysis in particular makes use of a bespoke alias analysis in order to determine whether values match or not;
the data race contract given in \cref{fig:contr_def_examples_c} would use this alias analysis to match the window handles of the function calls as well as
the buffer in the call to any memory access.
This alias analysis is critical: Inaccuracies will cause either no errors to be emitted,
e.g. if a memory access uses the buffer but is not matched by the alias analysis, or instead produce false positives,
e.g. if the window handle in the completion call is erroneously not matched to the communication call.

After the analyses finish, the post-processing pass resolves the contract formulas, performs error output, and hands off compilation to LLVM.

\subsection{CoVer-Dynamic}
\label{sec:cover-dynamic}

CoVer-Dynamic was built as a dynamic operating mode for the same contracts as those defined originally for use with CoVer-Static.
After CoVer-Static finishes its analysis, it may optionally instrument the program.
This adds callbacks to the module, which interface with functions defined in the CoVer-Dynamic runtime library that is linked to the executable once compilation finishes.
CoVer-Dynamic can then perform analysis at runtime using these callbacks.

More specifically, if dynamic analysis is requested, the compile wrapper of CoVer-Static will run an additional compile pass.
It embeds the parsed contracts of the contract manager into the resulting executable.
Further, callbacks to the CoVer-Dynamic runtime library are inserted for every memory access, and for function calls that have a contract attached or are mentioned in one.

The most important callback is the one made for function calls, as these can mark the beginning and end of contract requirements.
As CoVer-Dynamic, same as CoVer-Static, is intended to be programming-model agnostic, the function callback must be generic for any API call.
Simplified from \cite{orajiDynamicContractAnalysis2026}, the callback receives a pointer to the function called, the number of arguments used for that call,
and a variadic argument list describing the parameters of the function call.

\section{Implementation}
\label{sec:implementation}

While CoVer is meant for programming-model agnostic analysis,
it is currently dependent on the C/\CPP\ languages.
This section introduces the necessary enhancements made to support Fortran as well.

\subsection{Migrating the Contract Syntax to Fortran}
\label{sec:migrate_contracts}

A major part of CoVer is the contract language, which encodes the requirements to check during analysis.
It facilitates the independence from the programming model.
However, it was designed with the C/\CPP\ languages in mind,
suffering from some incompatibilities when porting to Fortran.
Thus, changes were necessary to the contracts themselves to allow analysis of Fortran code.
We will demonstrate the changes on the example contract given in \cref{fig:contr_def_examples_c};
the finished translation is given in \cref{fig:contr_def_examples_fort}.
\begin{figure}[tbp]
\begin{lstlisting}[language=fortran]
call Declare_Contract(MPI_Get, & 
   "PRE { call!(MPI_Init_f08) } &
  & POST { &
  &   no! (read!(0)) until! (call_tag!(rma_complete,$:7)), &
  &   no! (write!(0)) until! (call_tag!(rma_complete,$:7)) &
  & }"
)
call Declare_Contract(MPI_Win_fence, "TAGS { rma_complete(1) }")
call Declare_Contract(MPI_Win_unlock_all, "TAGS { rma_complete(0) }")
\end{lstlisting}
\caption{Example of contract definitions (Fortran).}
\label{fig:contr_def_examples_fort}
\end{figure}

\paragraph{Dereferencing / Address-Of Operator}
\label{sec:deref_addrof}

Consider the reading and writing operations given in the example contract in \cref{fig:contr_def_examples_c}.
These reference C-specific syntax for the dereferencing of a pointer, the \textquote{star (\texttt{*})} operator,
and the contract language also allows use of the \textquote{address-of (\texttt{\&})} operator.

While these are useful for describing common patterns in C code, they have no relevance for Fortran.
For example, when using C, programmers often use the address-of operator to pass parameters to function calls,
which are then dereferenced internally and written to as an analog of a pass-by-reference parameter.

However, Fortran instead employs the use of argument \emph{intent}.
A function definition specifies the intent of its arguments as \texttt{INTENT(\{IN,OUT,INOUT\})}.
These can then be used by the compiler to perform the corresponding code generation,
requiring no explicit action from the programmer calling the function.
Finally, this then causes issues with the static alias analysis,
as the function parameter arguments are not actually dereferenced / no \textquote{address-of} is being taken.
Thus, for our Fortran port, uses of \texttt{*/\&} are removed,
as they clash with both the Fortran language design and the alias analysis.

\paragraph{Fortran Generics}
\label{sec:fortran_generics}

While porting to Fortran, we were continuously testing using the MPI contracts,
as those were readily available as part of the predefined contracts of CoVer,
while also offering equivalent Fortran and C bindings.

However, functions referenced by name (such as \texttt{MPI\_Init} in \cref{fig:contr_def_examples_c}) were not being found during analysis.
The module-based Fortran MPI bindings (\texttt{use mpi}, \texttt{use mpi\_f08}) give the API function definitions as generics.
For the older \texttt{mpi} module, this does not change anything, but there is a critical difference for the \texttt{mpi\_f08} module:
The function \emph{implementation} within is named differently from the MPI call itself.
All MPI calls receive an additional \texttt{\_f08/\_f08ts} suffix, with the exact names specified in the MPI documentation \cite[p. 789]{messagepassinginterfaceforumMPIMessagePassingInterface2025}.
During compile time, the generic name is no longer available, even in debug information.
Thus, we require that the contracts either use the tag system for function calls, which can be matched directly,
or explicitly use the name of the function implementation.

\subsection{Adjustments for Static Analysis}
\label{sec:adjustments_static_analysis}

To allow for static analysis of Fortran code using CoVer,
we needed to add an alternative way to attach contracts, as well as a different alias analysis approach.
This section explains the specific changes made and the reasoning behind them.

\paragraph{Attaching Contracts to Functions in Fortran}
\label{sec:annot_with_funcdef}

The contract annotation system is based around the generic string attribute system of the \texttt{clang(++)} compiler.
When the code is translated to LLVM IR, the strings and corresponding function are stored together, making them accessible to the contract manager.
However, the closest analog we found for Fortran were compiler directives \cite{CompilerDirectivesSupported},
but these do not support string annotations as needed for the contracts.

Instead, for Fortran we add contracts using dummy \texttt{Declare\_Contract} functions,
which receive the function and the corresponding contract as parameters as done in \cref{fig:contr_def_examples_fort}.
Calls to this function correspond to a contract declaration,
and all such calls are deleted by the contract manager once contract parsing is complete.
These calls do not need to be inserted into the program code but can instead be added in new header files
to minimize impact on the application.
This method may also be easier to apply for future language ports.




\paragraph{Alias Analysis for Fortran}
\label{sec:alias_analysis_fortran}

Alias analysis (AA) is critical for error detection,
for example, when attempting to match memory accesses to function parameters as needed for the example contract in \cref{fig:contr_def_examples_fort}.

However, in our testing, both the built-in AA of CoVer-Static and the well-known SVF \cite{suiSVFInterproceduralStatic2016} tools were insufficient for the IR generated by \texttt{flang}.
No aliases could be detected at all, leading to rampant invalid results.
The built-in LLVM AA is unsuitable as it cannot be applied across functions.

Finally, we employed the LLVM Data Structure Analysis (DSA) \cite{lattnerDataStructureAnalysis}.
After porting the analysis to a newer LLVM version, it was able to find the necessary aliases
in the IR generated by \texttt{flang}.
CoVer now uses DSA by default for Fortran code, while relying on the existing AA implementation for C.

\subsection{Instrumentation for Dynamic Analysis}
\label{sec:instr_dynamic_analysis}

\begin{figure}[tbp]
    \begin{lstlisting}[language=llvm]
; The type shows how arrays contain not just a pointer, but additional metadata,
; e.g. size in bytes, rank (number of dimensions), length in each dimension, ^*\dots*^
C: %2 = alloca ptr, align 8
Fortran: %1 = alloca ^*\textbf{\{ ptr, i64, i32, i8, i8, i8, i8, ptr, [1 x i64] \}}*^, align 8

; When there are string parameters, additional length arguments are added
C: call i32 @MPI_Info_set(ptr %1, ptr %2, ptr %3) ; %2: "no_locks", %3: "true"
Fortran: call void @mpi_info_set_f08_(ptr %1, ptr %2, ptr %3, ptr %4, ^*\textbf{i64 8, i64 4}*^)
    \end{lstlisting}
  \caption{Overview of some additional Fortran metadata in the LLVM IR.}
  \label{fig:overview_fort_metadata}
\end{figure}

The dynamic analysis is largely decoupled from the rest of CoVer,
requiring only that the instrumentation is applied correctly.
However, some issues occurred with the function callbacks due to the way \texttt{flang} manages the Fortran metadata.
An overview of the metadata described here is given in \cref{fig:overview_fort_metadata}.

Fortran manages a lot of metadata itself for arrays such as their size or rank.
This means that this metadata must be carried along the array and into function calls,
causing the inserted callback to not receive the array, but instead a structure containing the array alongside its metadata.
This causes the dynamic analysis to be unable to match it to memory accesses, which instead reference the array itself.
To fix this, the instrumentation pass detects these structures,
and first extracts the array before generating the callback using it directly.

Similarly, \texttt{flang} manages string lengths as well.
However, instead of using a structure containing the string and its length, the length is added as an additional parameter \emph{after the expected parameters of the function};
e.g. \texttt{MPI\_Info\_set}, which expects 4 arguments (incl. the \texttt{ierror} parameter), two of which are strings, will instead receive 6 arguments at IR level.
This crashed for the current implementation of the instrumentation pass,
which attempted to extract nonexisting debug information for the additional arguments.
Fixing this required reducing the number of parsed arguments for dynamic analysis for each string argument.

\section{Evaluation}
\label{sec:Evaluation}

To prove the effectiveness of our Fortran port, we performed various tests regarding both the classification quality and overhead induced.
While the classification quality tests aim to uncover any possible accuracy regressions in CoVer when using Fortran instead of C,
the overhead analysis compares the execution runtime when using CoVer-Dynamic across languages.
We compare CoVer against MUST \cite{hilbrichMUSTScalableApproach2010} as a representative third-party dynamic correctness checker for differences in the transferability between C and Fortran
analysis accuracy or performance.
We were unable to do the same for a static one as to the best of our knowledge,
CoVer-Static is currently the only static correctness checker compatible with the \texttt{mpi\_f08} module.
The scripts, sources and results are available at \cite{orajiArtifactExtendingContract}.

The tests were run on the Lichtenberg II cluster with two Intel Xeon Platinum 9242 CPUs (96 cores) and 384 GB of RAM per node,
connected using Infiniband.

\subsection{Classification Quality}
\label{sec:classification_quality}

The classification quality tests use our recently introduced MPI-BugBench Fortran port \cite{jammerMPIBugBenchFrameworkAssessing2025,orajiExtendingMPICorrectness2025} (level 1).
This allows for a direct comparison of the effectiveness of CoVer on C and Fortran codes,
as the MPI-BugBench Fortran port contains an explicit mapping of Fortran tests to the respective C equivalents.

CoVer ran using MPICH 4.3.2, while MUST was tested with both (a modified) MPICH 4.3.2 and OpenMPI 5.0.5.
While CoVer works equally well with any implementation (it is decoupled from any programming model entirely),
MUST requires a workaround to analyse code using the \texttt{mpi\_f08} module; it requires the Fortran MPI implementation to call an underlying, interceptable C MPI interface.
OpenMPI does not do this, and while MPICH normally calls the PMPI C functions (thus not allowing interception), this can be patched in the MPICH source code.
Thus, MUST requires a modified MPICH(-based) MPI implementation to perform analysis on \texttt{mpi\_f08} code.

Furthermore, MUST deadlocks if checking for both MPI RMA and P2P data races when using an MPICH implementation.
As the default configuration in MPI-BugBench does not make use of RMA detection,
this was not caught during the evaluation of the MPI-BugBench Fortran port \cite{orajiExtendingMPICorrectness2025}.
We therefore tested MUST using both MPICH and OpenMPI, using MPICH for Fortran but without RMA checks and OpenMPI for C but including RMA checks as well.

In total, we ran the evaluation for all Fortran test cases of the MPI-BugBench level 1 suite, as well as all equivalent C tests (222 tests total), for CoVer-Static, CoVer-Dynamic, MUST/MPICH (C+Fortran) and MUST/OpenMPI (C).
The tests were grouped into true positives (TP), true negatives (TN), false positives (TP) and false negatives (TN)\footnote{NC-TP and NC-FP from hybrid static/dynamic analysis \cite{orajiCouplingStaticDynamic2025} treated as TP and FP.}.
We also calculated the accuracy as $\frac{TP+TN}{\#tests}$.

To avoid comparing trivial FNs with unsupported error classes, we are focusing on those currently supported as described in \cite{orajiDynamicContractAnalysis2026} (Missing init-/finalization, data races, handle lifecycle, RMA errors).
This allows for a more direct comparison of possible regressions due to the Fortran port.
We did, however, run each tool on \emph{all} tests to catch possible FPs, though none occurred.

\begin{table}[tbp]
    \centering
    \caption{Results for MPI-BugBench level 1, supported tests from \cite{orajiDynamicContractAnalysis2026}.}
\begin{tabularx}{\textwidth}{lXcccccXcccccXccccc}
\toprule
\multirow{3}{*}{Tool} & & \multicolumn{5}{c}{Fortran} & & \multicolumn{5}{c}{C} & & \multicolumn{5}{c}{Difference} \\
\cmidrule(lr){3-7}\cmidrule(lr){9-13}\cmidrule(lr){15-19}
 & & TP & TN & FN & TO & Acc & & TP & TN & FN & TO & Acc & & TP & TN & FN & TO & Acc \\
\midrule
CoVer-Dynamic &  & 22 & \cellcolor{green!20}23 & 2 & 0 & 0.96 &  & 22 & \cellcolor{green!20}23 & 2 & 0 & 0.96 &  & $+0$ & $+0$ & $+0$ & $+0$ & $+0.00$ \\
CoVer-Static &  & \cellcolor{green!20}24 & \cellcolor{green!20}23 & \cellcolor{green!20}0 & 0 & \cellcolor{green!20}1.00 &  & \cellcolor{green!20}24 & \cellcolor{green!20}23 & \cellcolor{green!20}0 & 0 & \cellcolor{green!20}1.00 &  & $+0$ & $+0$ & $+0$ & $+0$ & $+0.00$ \\
MUST &  & 17 & \cellcolor{green!20}23 & 6 & \cellcolor{red!20}1 & 0.87 &  & 21 & \cellcolor{green!20}23 & 2 & \cellcolor{red!20}1 & 0.96 &  & \cellcolor{red!20}$-4$ & $+0$ & \cellcolor{red!20}$+4$ & $+0$ & \cellcolor{red!20}$-0.09$ \\
\bottomrule
\end{tabularx}
\begin{flushleft}
\scriptsize
\hspace*{0.5em} TP\textuparrow: True Positive, TN\textuparrow: True Negative, FN\textdownarrow: False Negative, A\textuparrow: Accuracy, TO\textdownarrow: Timeout (30s)\\
\hspace*{0.5em} False positives omitted as none occurred. \textuparrow/\textdownarrow: Higher/Lower is better.\\
\hspace*{0.5em} Best result highlighted in green. TOs and accuracy regressions marked red.
\end{flushleft}
    \label{fig:results_classification_quality}
\end{table}
The results for the classification quality tests can be seen in \cref{fig:results_classification_quality}.
As we are focusing on regression testing, the table section on differences of the Fortran results to C is most interesting:
For CoVer, the results are exactly the same across languages.
Thus, there are no accuracy regressions for the tests given in MPI-BugBench;
the port perfectly preserves CoVer's capabilities.

The same cannot be said for MUST: Due to the aforementioned limitations with the analysis of RMA and non-RMA code, we were forced to disable some checks for MUST to function correctly.
This caused MUST to be unable to detect \emph{any} RMA races in the Fortran tests, lowering its accuracy by 9\%.

Further, during the evaluation we noticed an issue with one test case.
Both CoVer-Static and CoVer-Dynamic reported an error on a test case marked as containing no errors by the MPI-BugBench project.
The test case indeed turned out to be invalid: It was waiting on the same request twice.
We fixed the error in the test case generation and published a fix upstream, which has since been merged by the maintainers of the project \cite{orajiFixUnmatchedWait2026}.

\subsection{Overhead Analysis}
\label{sec:overhead_analysis}

For the overhead analysis, we wanted to perform again a direct comparison of C and Fortran code, similar to the classification quality tests.
We therefore need proxy applications which are available in both C and Fortran versions.
For this, we used the C and Fortran versions of miniWeather \cite{normanMiniWeather2020} (\texttt{mpi} Fortran Module).
To add another point of comparison, we ported the PRK Stencil \cite{vanderwijngaartParallelResearchKernels2014} MPI RMA code to Fortran (\texttt{mpi\_f08} Fortran Module) as well.

Using these two proxy applications with both Fortran and C, we ran a baseline execution, CoVer-Dynamic and MUST five times each to measure the overhead caused by each tool.
Deviations were at most 10-20\%, which is negligible considering that the runtimes of different configurations differ by at least 2x.

\newcommand{\langDiffTable}[2]{
\multirow{6}{0.5cm}{\rotatebox{90}{#2}}& MUST &  48 & \pgfplotstablegetelem{0}{must_fort}\of{#1}\pgfmathprintnumber[fixed,precision=2]{\pgfplotsretval}s & \pgfplotstablegetelem{0}{ratio_must}\of{#1}\pgfmathprintnumber[fixed,precision=2,zerofill]{\pgfplotsretval}x\\
&      &  96 & \pgfplotstablegetelem{1}{must_fort}\of{#1}\pgfmathprintnumber[fixed,precision=2]{\pgfplotsretval}s & \pgfplotstablegetelem{1}{ratio_must}\of{#1}\pgfmathprintnumber[fixed,precision=2,zerofill]{\pgfplotsretval}x\\
&      & 192 & \pgfplotstablegetelem{2}{must_fort}\of{#1}\pgfmathprintnumber[fixed,precision=2]{\pgfplotsretval}s & \pgfplotstablegetelem{2}{ratio_must}\of{#1}\pgfmathprintnumber[fixed,precision=2,zerofill]{\pgfplotsretval}x\\
& CoVer &  48 & \pgfplotstablegetelem{0}{cover_fort}\of{#1}\pgfmathprintnumber[fixed,precision=2]{\pgfplotsretval}s & \pgfplotstablegetelem{0}{ratio_cover}\of{#1}\pgfmathprintnumber[fixed,precision=2,zerofill]{\pgfplotsretval}x\\
&       &  96 & \pgfplotstablegetelem{1}{cover_fort}\of{#1}\pgfmathprintnumber[fixed,precision=2]{\pgfplotsretval}s & \pgfplotstablegetelem{1}{ratio_cover}\of{#1}\pgfmathprintnumber[fixed,precision=2,zerofill]{\pgfplotsretval}x\\
&       & 192 & \pgfplotstablegetelem{2}{cover_fort}\of{#1}\pgfmathprintnumber[fixed,precision=2]{\pgfplotsretval}s & \pgfplotstablegetelem{2}{ratio_cover}\of{#1}\pgfmathprintnumber[fixed,precision=2,zerofill]{\pgfplotsretval}x\\
}
\newcommand{\parseTableLangDiff}[2]{
    \pgfplotstableread[col sep=semicolon]{#1}#2
    \pgfplotstablecreatecol[
        create col/expr={\thisrow{must_fort}/\thisrow{must_c}}
    ]{ratio_must}#2
    \pgfplotstablecreatecol[
        create col/expr={\thisrow{cover_fort}/\thisrow{cover_c}}
    ]{ratio_cover}#2
}
\parseTableLangDiff{images/data/miniWeather_runtimes.csv}{\miniWeatherTable}
\parseTableLangDiff{images/data/stencil_runtimes.csv}{\stencilTable}

\begin{figure}[tbp]
    \centering
    \newlength{\overheadPanelHeight}
    \setlength{\overheadPanelHeight}{3.7cm}
    \begin{subfigure}[b]{0.58\textwidth}
        \centering
        \begin{minipage}[c][\overheadPanelHeight][c]{\textwidth}
            \centering
            \includegraphics[width=\textwidth]{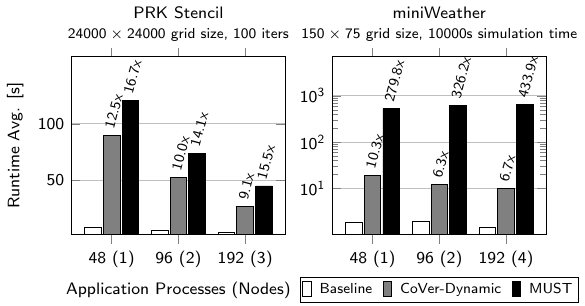}
        \end{minipage}
        \caption{Fortran runtimes and factor over baseline.}
    \end{subfigure}
    \hfill
    \begin{subfigure}[b]{0.41\textwidth}
        \centering
        \begin{minipage}[c][\overheadPanelHeight][c]{\textwidth}
            \centering
            \resizebox{\textwidth}{!}{
            \begin{tabular}{clccc}
                \toprule
                & Tool & Procs. & Runtime (Fort) & Factor over C\\
                \midrule
                \langDiffTable{\stencilTable}{PRK Stencil}%
                \midrule
                \langDiffTable{\miniWeatherTable}{miniWeather}%
                \bottomrule
            \end{tabular}
            }
        \end{minipage}
        \caption{Runtimes and factor over C}
    \end{subfigure}
    \caption{Results of the overhead analysis.}
    \label{fig:results_overhead_analysis}
\end{figure}
The results of the performance evaluation are shown in \cref{fig:results_overhead_analysis}.
We see that the improved overhead of CoVer-Dynamic compared to MUST mirrors the results on the C proxy applications measured in \cite{orajiDynamicContractAnalysis2026}.

However, we also note significant performance regressions compared to the C baseline.
While MUST is relatively consistent regarding overhead for both C and Fortran executions,
CoVer-Dynamic on Fortran increases runtimes from +30\% up to +130\% in the worst case.

The cause of this discrepancy lies in the increased amount of instrumented memory accesses.
While the C and Fortran codes of both applications are largely equivalent,
CoVer-Dynamic struggles to differentiate between irrelevant accesses to variables and their metadata.
As mentioned in \cref{sec:instr_dynamic_analysis}, \texttt{flang}-generated IR often carries additional metadata for variables.
CoVer-Dynamic inserts additional instrumentation when it is accessed, causing significant slowdown compared to C.
We have implemented multiple methods to reduce the number of instrumented metadata accesses,
though currently the number of instrumented calls due to memory accesses is still around 2x to that of C code.

\section{Related Work}
\label{sec:related_work}

Much work has gone into automatic detection of parallel programming errors.
Tools such as the MPI-Checker \cite{drosteMPIcheckerStaticAnalysis2015} and the SPMD IR \cite{burakSPMDIRUnifying2025}
allow for static analysis thereof, though are limited to C only.

MUST \cite{hilbrichMUSTScalableApproach2010}, RMASanitizer \cite{schwitanskiRMASanitizerGeneralizedRuntime2024} and PARCOACH \cite{saillardPARCOACHCombiningStatic2014}
all support Fortran, but with significant limitations.
They each only support the older \texttt{mpif.h} or \texttt{use mpi} modules, where the former has already been deprecated and the latter strongly discouraged in favor of the new \texttt{use mpi\_f08} module.
It is currently not possible to perform analyses of \texttt{mpi\_f08} code with PARCOACH, and both MUST and RMASanitizer depend on a modified MPICH implementation to work.

Fortran-native approaches, such as the dynamic analysis tool MPI-CHECK \cite{lueckeMPICHECKToolChecking2003},
are in turn is limited to Fortran only, and, compared to the contract-based CoVer, cannot check arbitrary parallel programming models or error classes.
Finally, similar to PARCOACH, MPI-CHECK only checks Fortran 90 code.

To the best of our knowledge, CoVer-Static is the first static analysis tool for Fortran codes which use the \texttt{mpi\_f08} module.
While dynamic analysis is possible using workarounds with MUST, CoVer-Dynamic does not depend on a specific MPI library or modifications thereof.
It is compatible with older standards as well, while keeping the generic, contract-based nature of the tool intact.

\section{Conclusion}
\label{sec:conclusion}

Programming distributed-memory clusters requires parallel programming models such as MPI.
While powerful, these models are quite error-prone.
Many correctness checkers have been created to aid developers,
though often limited to C codes, specific library implementations incompatible with the cluster or program to be debugged,
or they do not check for the required error class.

Previously, we demonstrated how CoVer \cite{orajiVerifyingMPIAPI2026}, a generic verification framework for parallel programming models,
allows for both static and dynamic analysis of arbitrary error classes.
However, it too was limited to the analysis of C/\CPP\ codes, contradicting the intended generic nature of the tool.

Thus, this work further builds on the applicability of CoVer:
By extending support to Fortran, CoVer can now be used in more development environments.
We have amended the language-specific parts of the contract definitions,
and modified the analyses to allow for analysis of programs written in either language.

Our evaluation shows that the port preserves the analysis accuracy of CoVer;
we found no accuracy regressions but instead uncovered a bug in the MPI-BugBench testing framework \cite{orajiFixUnmatchedWait2026} itself,
proving the effectiveness of our tool.
Our port, while causing increased overhead compared to C, still causes significantly less runtime overhead than MUST \cite{hilbrichMUSTScalableApproach2010},
while retaining feature parity.

In future work the additional instrumentation of Fortran metadata should be reduced.
Further, since the contracts were almost directly transferrable to Fortran,
additional language ports can be explored.

\begin{credits}
\subsubsection{\ackname}
\NHRText
We thank Alexander Hück for his helpful feedback.

\subsubsection{\discintname}
The authors have no competing interests to declare that are
relevant to the content of this article.
\end{credits}
%
%
%
%
\bibliographystyle{splncs04}
\bibliography{bibliography}

\end{document}